\documentstyle[prd,aps]{revtex}
\begin{document}
\title{The classical essence of black hole radiation} 
\author{M.~Nouri-Zonoz\thanks{Electronic address:~nouri@iucaa.ernet.in} and 
T.~Padmanabhan\thanks{Electronic address:~paddy@iucaa.ernet.in} }
\address{IUCAA, Post Bag 4, Ganeshkhind, Pune 411 007, INDIA.}


\maketitle

\begin{abstract}
We show that the mathematics of Hawking process 
can be interpreted classically as the Fourier analysis of an 
exponentially redshifted wave mode which scatters off the 
black hole and travels to infinity at late times. 
We use this method to derive the Planckian power spectrum for 
Schwarzchild, Reissner-Nordstrom and Kerr black holes.

\end{abstract}


\section{Introduction}
The study of black holes from astrophysical point of view and by 
astronomers has 
 blossomed in the last decade because of the dramatic increase
in the number of black hole candidates from the sole 
 candidate (Cygnus X-1) some 25 years ago. This in turn 
requires a deeper familiarity with black hole physics and especially
with {\it black hole radiation}, for astronomers and classical 
relativists. Hawking in his original work on the black hole 
radiation ~(Ref.[4]) has used a quantum field theoretical 
approach to arrive at this radiation.
In the next few sections we will 
describe the classical essence of this radiation in a language
 which is free from the usual quantum
field theoretic tools and therefore more suitable for the 
astronomers and relativists. 
Our derivation of Hawking
radiation will also establish the  close connection between the black 
hole radiation and the existence of an event horizon.

\section{Schwarzschild black hole}\label{sec:Schwaz} 
We start with the simplest case of the  
one parameter family of blackholes, namely the Schwarzschild black hole,
which was previously discussed in Ref.[6]. Consider a
radial light ray in the Schwarzschild spacetime which propagates
from $r_{in} = 2M + \epsilon $ at $t=t_{in}$ to the 
event ${\cal P}(t,r)$
where $r \gg 2M$ and $ \epsilon \ll 2M$.
The trajectory can be found using the fact that for light rays 
$${\rm d}s^2=1 - {2M \over r}{\rm d}t^2 - {1\over {1 - {2M \over r}}}
{\rm d}r^2 - r^2 {\rm d}\Omega^2 = 0,$$ 
for radial light rays, ${\rm d}\theta = {\rm d}\phi =0$, we have  
$${{\rm d}r \over {\rm d}t} ={1 - {2M \over r}},\eqno(1)$$ 
from which the trajectory with the requierd initial condition, is
$$r=r_{in} - 2M{\rm ln}({r-2M \over 2M}) + 2M{\rm ln}
({r_{in} - 2M \over 2M})+ t - t_{in}
 \cong t - t_{in} + 2M {\rm ln} ({\epsilon \over 2M})\eqno(2)$$  
where the last equality uses $r\gg 2M$ , $\epsilon \ll 2M$.
The frequency of a wave will be redshifted  as it propagates on this 
trajectory. This redshift is basically due to the fact that the
frequency is measured in terms of the proper time $\tau$, which flows 
differently at different points of a stationary spacetime according to the 
following relation:
$$\tau ={1\over c}\sqrt{g_{00}} x^0. \eqno(3)$$  
The frequency $\Omega$ at $r\gg 2M$ will be related to the 
frequency $\Omega_{in}$ at $r =2M + \epsilon$ by
$$\Omega \cong \Omega_{in}[g_{00}(r=2M+\epsilon)]^{1/2} \cong
\Omega_{in} \left( \epsilon \over 2M \right)^{1/2} = \Omega_{in}
{\rm exp}\left( -{{t - t_{in} - r}\over 4M}\right) \eqno(4)$$
If the wave packet, $\Phi(r,t)\propto {\rm exp}({\it i}\theta(t,r))$, 
centered on this null ray has a phase $\theta (r,t)$, then the 
instantaneous frequency is related to the phase by 
$(\partial \theta / \partial t) = \Omega$. Integrating (4) with
respect to $t$, we find the relevant wave mode to be
$$\Phi(t,r) \propto {\rm exp} {\it i}\int \Omega{\rm d} t \propto 
{\rm exp} \left[ -4M{\it i}\Omega_{in} {\rm exp} \left(-{{t - t_{in} - r}
\over 4M}\right)\right]\eqno(5)$$
(This form of the wave  can also be obtained by directly 
integrating the wave equation 
in Schwarzschild space with appropriate boundary conditions).
Equation (4) shows that, despite being in a static spacetime, the 
frequency of the wave (measured by an observer at fixed $r\gg 2M$)
depends nontrivially on $t$, for a fixed $t_{in}$ and $\epsilon$.
Such an observer will not see a monochromatic radiation.  
Therefore an observer using the time coordinate $t$ will Fourier decompose 
these modes with respect to the frequency $\omega$ defined using $t$ as:
$$\Phi(t,r)={1\over 2\pi} \int^{\infty}_{-\infty} f(\omega) 
e^{-i\omega t} {\rm d}\omega \eqno(6)$$
where
$$f(\omega)= \int^{\infty}_\infty \Phi(t,r) e^{i\omega t}{\rm d}t \propto
\int^{\infty}_0 x^{-4iM\omega - 1} {\rm exp}(-4Mix \Omega_{in})
{\rm d}x \eqno(7) $$
and $x={\rm exp} \left(\; [-t + t_{in} + r]/ 4M \right)$.
To evaluate the above integral we rotate the contour to the imaginary 
axis, i.e. $x \rightarrow y=ix$,
$$f(\omega) \propto e^{-2\pi M \omega}\int^{i\infty}_0 Y^{z-1} e^{-Y}
dY \eqno(8)$$
where $z=-4 i M \omega$ and $Y=-4 M\Omega_{in}y$.
Using the fact that the integral in the right hand side of the above
relation is one of the  representations of Gamma function we get 
the corresponding power spectrum to be
$$|f(\omega)|^2 \propto ({\rm exp}(8\pi M\omega) -1)^{-1}\eqno(9)$$
where we have used the fact that $|\Gamma (ix)|^2 = {(\pi /
x {\rm sinh} \pi x)}$.
In terms of the  conventional units the above relation becomes
$$|f(\omega)|^2 \propto ({\rm exp}({8\pi G M \omega \over c^3}) -1)^{-1}
\equiv (exp({\omega \over \omega_0}) -1)^{-1}\eqno(10)$$
where $$\omega_0 = {c^3\over 8\pi G M}\eqno(11).$$ 
As one can see no $\hbar$ appears in the above analysis and 
$\omega_0$ can be thought of as the characteristic frequency of the
problem by a radio astronomer who thinks in terms of frequency. 
On the other hand an X-ray or a $\gamma$-ray 
astronomer -who thinks in terms of photons- will 
introduce the energy $E=\hbar \omega$ into the above relation 
in the following form:
$$|f(\omega)|^2 \propto (exp({\hbar \omega \over \hbar \omega_0}) -1)^{-1}
\equiv (exp({E \over k_{_B}T}) -1)^{-1}\eqno(12)$$
which shows that the corresponding  power spectrum is Planckian at 
temperature
$$T = {\hbar c^3 \over 8\pi G M k_{_B}} .\eqno(13)$$
\section{Reissner-Nordstrom balck hole}\label{sec:R-N}
The same approach can be used 
to study the radiation in the space of static charged black holes which are
charcterized by two parameters M and Q.    
The equation governing the  outgoing null radial geodesics in R-N 
spacetime has the following form
$${{\rm d}r \over {\rm d}t} = 1 - {2M \over r} + {Q^2 \over r^2}\eqno(14)$$
In terms of the conventional units the above equation will take
the following form
$${{\rm d}r \over {\rm d}t} = c - {2M \over r}({G\over c}) + {Q^2 \over r^2}
({G\hbar\over c^2}).\eqno(15) $$
The event horizon of the Reissner-Nordstrom balck hole is 
at $r_{+} = M + (M^2 - Q^2)^{1/2}$. 
Considering a light ray propagating from $r_{in} = r_{+} +
\epsilon$ at $t=t_{in}$ to the event ${\cal P}(t,r)$
where $r \gg r_{+} $ and $ \epsilon \ll r_{+} $ we will find the
trajectory in the follwing form
$$r \cong t - t_{in} + {{r_{+}}^2\over 2(M^2 - Q^2)^{1/2}}{\rm ln} \epsilon
\eqno(16)$$
The redshifted frequency $\omega$ will be related to the frequency
at $r = r_{+} + \epsilon$ by
$$\Omega \cong \Omega_{in}[g_{00}(r=r_{+}+\epsilon)]^{1/2} \cong
\omega_{in} \left( \epsilon \over 2M \right)^{1/2} = \omega_{in}
{\rm exp}\left( -{{t - t_{in} - r}\over {(M+(M^2-Q^2)^{1/2})^2 \over 
(M^2-Q^2)^{1/2}}}\right) \eqno(17)$$
Now if we repeat the analysis of the Schwarzschild case for R-N
spacetime, in exactly the same way, we find that the corresponding
power spectrum for a wave packet which has scattered off the R-N
black hole and travelled to infinity at late times has the
following Planckian form
$$|f(\omega)|^2 \propto \left( {\rm exp}\left[{2\pi[M+(M^2-Q^2)^{1/2}]^2 \over
(M^2-Q^2)^{1/2}} \right]\omega - 1 \right)^{-1}\eqno(18)$$
at temperature $T = {(M^2-Q^2)^{1/2}\over
2\pi[M+(M^2-Q^2)^{1/2}]^{2}}$
which is the standard result and reduces to that of the Schwarzschild case 
when $Q=0$.  
\section{Hawking radiation of a Kerr black hole}\label{sec:Kerr}
In applying the approach of the last two sections to the radiation
of Kerr black holes 
we should be more careful because- unlike the Schwarzschild and
R-N black holes- the event horizon and 
infinte redshift surface do not coincide. We will see that 
in this case the infinte redshift surface acts as a boundary for the 
outgoing null geodesics originating from inside the ergosphere, on which
we should be concerned about the continuty problem.
In Kerr spacetime the principal null congruences play the same role
as the radial null geodesics in Schwarzschild and R-N spacetimes,
so we consider them in our derivation of the Hawking radiation
by Kerr black holes. The equation governing the principal null
congruences ($\theta$ = const.) is given by
$${{\rm d}r \over {\rm d}t} = 1 - {2M\over r} + {a^2 \over r^2}
\eqno(19)$$
If we restrict our attention to the case $a^2 < M^2$, the above 
equation can be integrated to give
$$t=r+\left( M + {M^2\over (M^2 - a^2)^{1/2}}\right){\rm ln}|r-r_+|
+\left( M - {M^2\over (M^2 - a^2)^{1/2}}\right){\rm ln}|r-r_-|\eqno(20)$$
where
$$r_{\pm} = M \pm (M^2 - a^2)^{1/2}\eqno(21)$$
are the event horizons of the Kerr metric. 
Now as in the previous 
sections we consider a light ray propagating from point
 $r_+ + \epsilon$  at $t=t_{in}$ to the 
event ${\cal P}(r,t)$ 
where $r,t \gg M $ and $\epsilon \ll M$.
Starting from a point very close to the outer event 
horizon ($r_+ +\epsilon$) the trajectory would have the following form
$$r \cong t - t_{in} + \left(M+{M^2\over 
(M^2 - a^2)^{1/2}}\right){\rm ln} \epsilon \eqno(22)$$

The frequency $\Omega$ at $r$ will be related to the 
frequency $\Omega_{in}$ of a light ray emitted by a {\it locally
nonrotating observer} (Ref.[1]) at 
$r =r_{+} + \epsilon$ (inside the 
ergosphere) by (see appendix A)
$$\Omega = \Omega_{in}{\left( g_{00} - {g^2_{03}/ g_{33}}\right)^{1/2}
\over (1+({g_{03}/ g_{00}})a {\rm sin}^2\theta )} \propto \Omega_{in} 
\epsilon^{1/2} = \Omega_{in}
{\rm exp} \left( -{(t - t_{in} - r)(M^2-a^2)^{1/2}\over 2(M^2
+M(M^2-a^2)^{1/2}) }\right)\eqno(23)$$
repeating the procedure of the last two sections to the above 
redshifted frequncy we find the following power spectrum for a
wave packet scattered off the Kerr black hole at late times

$$|f(\omega)|^2 \propto \left( {{\rm exp} 
\left[ 4\pi[M^2+M(M^2-a^2)^{1/2}] \over
(M^2-a^2)^{1/2}\right]\omega} - 1 \right)^{-1}\eqno(24)$$
which is Planckian at temperature

$$T = {(M^2-a^2)^{1/2}\over
4\pi[M^2+M(M^2-a^2)^{1/2}]}.\eqno(25)$$
which is again the standard result (Ref.[2]) and reduces to (13) for $a=0$. 

\section{Discussion}\label{sec:cnclsns}
In this letter we gave a simple derivation of balck hole radiation which
strips the Hawking process to its bare bones and establishes the following
two facts: 
(i) The key input which leads to the Planckian spectrum is the 
exponential redshift given by equations (4,17 \& 23) of modes which scatter 
off the black hole and travel to infinity at late times, which in turn 
requires the existence of an event horizon. It is well known 
that frequencies of outgoing waves at late times in black hole 
evaporation correspond to super planckian energies of the ingoing
modes near the horizon. One might ask where do
the ingoing modes corresponding to the outgoing modes come from?. 
This where the quantum field theory plays its role
in the black hole radiation. According to quantum field theory vacuum 
is a dynamical entity and space is nowhere free of vacuum fluctuations.
The vacuum field fluctuations can be thought of as a superposition
of ingoing and outgoing modes. A {\it collapsing star}
will introduce a mismatch between these virtual modes causing
the appearance of a real particle at infinity. The calculation shows that 
the energy carried by the radiation is extracted from the black 
hole (Ref.[4]). What we have done is to mimic the essence of this 
process by considering a classical mode propagating from near
event horizon to infinity.  
(ii) The analysis given in the previous sections is entirely 
classical and no $\hbar $ appears anywhere.
The mathematics of Hawking evaporation is puerly classical and lies in
the Fourier transform of an exponentially redshifted 
wave mode (for a more detailed discussion of classical versus quantum 
features see ref. [7]).
\section*{Appendix A : Gravitational redshift by a Kerr black hole}
In this appendix we derive the  gravitational redshift
of a light ray emmitted from inside the ergosphere and received by
a Lorentzian observer at infinity (as given by equation (17) of the text).
The general relation for the redshift between a source and an observer
located at events ${\cal P}_1$ and ${\cal P}_2$ 
in an stationary spacetime, is given by (Ref[8])
$${\omega_{\;_{{\cal P}_1}} \over \omega_{\;_{{\cal P}_2}}} = {(k_a u^a)_
{{\cal P}_1}\over (k_a u^a)_{{\cal P}_2}}\eqno(A1)$$
where $k^a$ is the wave vector and $u^a$s are the 4-velocities of 
the source and the observer. The numerator and denominator are 
evaluated at the events ${\cal P}_1$ and ${\cal P}_2$ respectively. One
should note that the null geodesic 
(or equivalently its tangent vector) joining the source and the observer 
should be continuous over the boundary which in this case is the 
infinte redshift surface. The principal null congrunces we are considering
here, indeed satisfy this condition.
Since there are no static observers inside 
the ergosphere we choose as our source the {\it locally nonrotating 
observer} ~(Ref.[1]) whose angular velocity in Boyer-Lindquist 
coordinates is given by ~(Ref[5])
$$\Omega= -{g_{03}\over g_{33}}={2Mra \over (r^2 + a^2)^2 - \Delta a^2 
{\rm sin}^2\theta}\eqno(A2) $$
so the 4-velocities of the source and the static Lorentzian observer
are given by (Ref[5])
$$u^a|_S = {1\over (g_{00} + 2\Omega g_{03} + \Omega^2 g_{33})^{1/2}}
(1, 0, 0, \Omega)\;\;\ \& \;\; u^a|_\infty
=(1/\sqrt{g_{00}}, 0, 0, 0)\eqno(A3)$$
Substituting (A2) and (A3) in (A1) we have
$$\omega |_\infty = \omega|_S \left({k_0 |_{\infty}\over 
(k_0 u^0 + k_3 u^3)|_S}\right) $$
Using the fact that the frequency measured with respect to
the coordinate time, $k_0$, is constant and that $k_3 / k_0 =-a \rm sin^2 
\theta$  for the principal null congruences ~(Ref[3]) we have
$$\omega |_{\infty} = \omega|_S \left({1 \over 
(u^0 - a \rm sin^2 \theta u^3)|_S}\right)\eqno(A4)$$
Now Substituting from (A2) and (A3) in (A4) we obtain the following 
result 
$$\omega |_{\infty} = \omega|_S {\left( g_{00} - {g^2_{03}
/g_{33}}\right)^{1/2}
\over (1+({g_{03}/ g_{00}})a {\rm sin}^2\theta )}\eqno(A5)$$ 
which is the relation used in the text.\\  



\end{document}